\documentclass[aps,prl,reprint,groupedaddress]{revtex4-1}

\usepackage{graphicx}
\usepackage{subfigure}

\usepackage{verbatim}

\begin{document}



\title{Network-like propagation of cell-level stress in random foams}

\author{Myfanwy~E.~Evans}
\email{myfanwy.e.evans@physik.uni-erlangen.de}
\author{Andrew~M.~Kraynik}
\author{Douglas~A.~Reinelt}
\author{Klaus~Mecke}
\author{Gerd~E.~Schr\"oder-Turk}
\affiliation{\mbox{Theoretische Physik, Friedrich-Alexander~Universit\"at Erlangen-N\"urnberg, Staudtstr.~7B, 91058 Erlangen, Germany}}
\affiliation{\mbox{Department of Mathematics, Southern Methodist University, Dallas, Texas 75275-0156}}

\date{\today}

\begin{abstract}
Quasistatic simple shearing flow of random monodisperse soap froth is investigated by analyzing Surface Evolver simulations of spatially periodic foams. Elastic-plastic behavior is caused by irreversible topological rearrangements (T1s) that occur when Plateau's laws are violated; the first T1s occur at the elastic limit and at large strains frequent cascades of T1s, composed of one or more individual T1s, sustain the yield-stress plateau. The stress and shape anisotropy of individual cells is quantified by $Q$, a scalar measure derived from the interface tensor that gauges each cell's contribution to the global stress. During each T1 cascade, the connected set of cells with decreasing $Q$, called the \textit{stress release domain}, is network-like and highly non-local. Geometrically, the network-like nature of the stress release domains is corroborated through morphological analysis using the Euler characteristic. The stress release domain is distinctly different from the set of cells that change topology during a T1 cascade. Our results highlight the unique rheological behavior of foams, where complex large-scale cooperative rearrangements of foam cells are observed as a consequence of distinctly local events. 

\end{abstract}


\maketitle


Foams are complex fluids with a shear modulus and a yield stress that are linked to shear-induced deformations of jammed nonspherical bubbles~\cite{Kraynik:88,HohlerCohenAddad:05}. Shaving foam furnishes a common but illustrative rheological experience; it is a soft solid at low stress yet flows at high stress. The defining mechanisms behind yield stress fluids are predominant under quasistatic conditions where the deformation rate is small and viscous forces are negligible. The shear modulus and yield stress characterize the elastic-plastic response, and both increase as the volume fraction of gas increases, reaching a maximum in the dry foam limit where the liquid volume fraction approaches zero~\cite{Princen:85,Princen:86}. Remarkably, foam is stiffest and strongest when it contains mostly gas.

\begin{figure}
\includegraphics[height = 0.47\columnwidth]{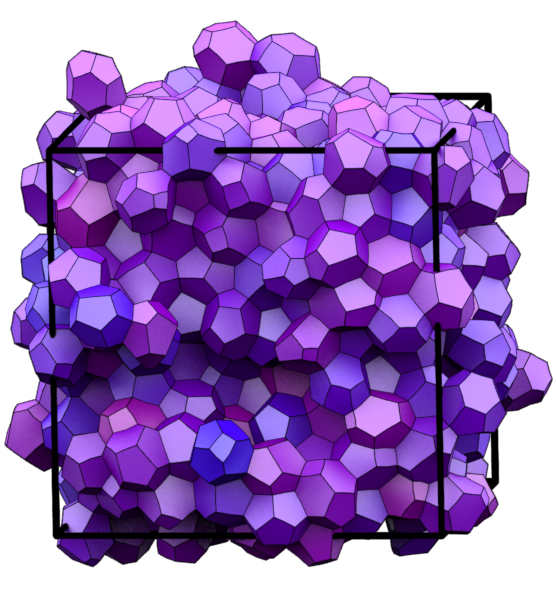}
\includegraphics[height = 0.47\columnwidth]{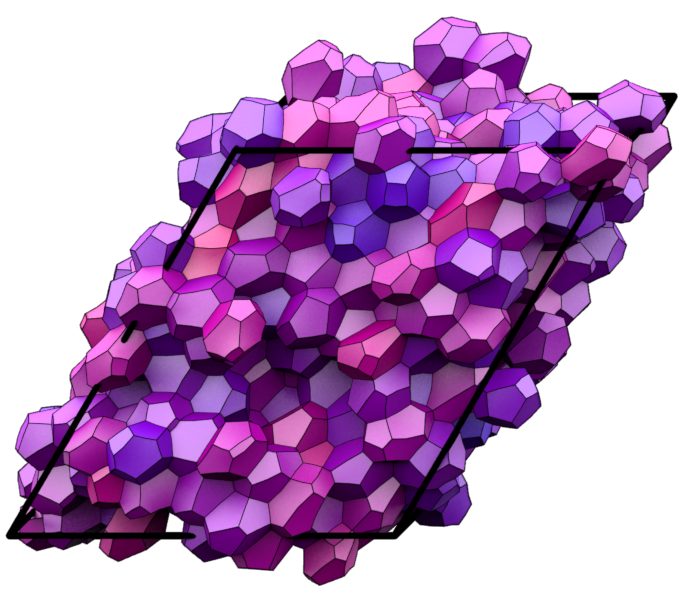}
\caption{(Color online) A dry liquid foam with 512 cells per periodic unit cell at strains of $0$ and 0.6 respectively. The cells are colored by $Q$, a scalar measure of cell distortion, where blue through red symbolizes values 0 through 1. The average $Q$ over all cells are 0.244 and 0.425 respectively.}
\label{Colour_by_Q}
\end{figure}

\begin{figure}
\includegraphics[width = 0.89\linewidth]{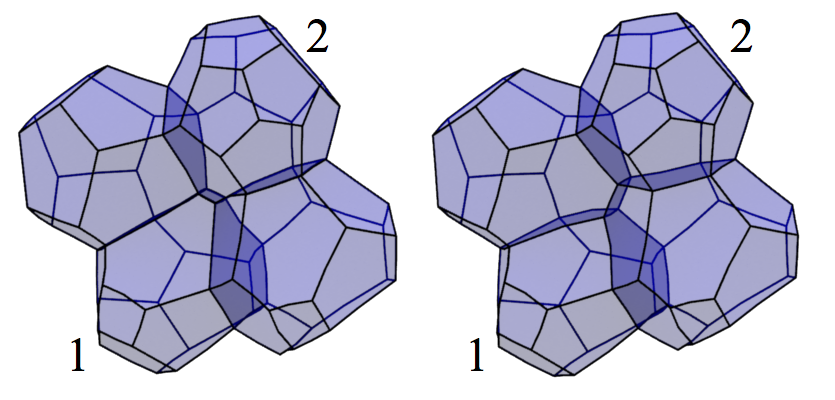}
\caption{(Color online) Illustration of a T1 topological transition in three dimensions, during which cells 1 and 2 lose contact and the other two cells become neighbors.}
\label{Topo}
\end{figure}

Consider a dry soap froth subjected to quasistatic shear (Figure~\ref{Colour_by_Q}). The idealized microstructure consists of a network of surfaces that divide space into polyhedral cells and represent thin liquid films with negligible thickness. Dry foams in equilibrium satisfy Plateau’s laws, which state that three films meet along each edge at $120^\circ$, and four edges meet at each vertex~\cite{Plateau}. Plateau's laws govern the local geometry of the equilibrium film network as a consequence of surface energy (area) minimization, and therefore, determine the rheology of the soap froth. The response of foam to deformation is elastic until Plateau's laws are violated; typically, a cell edge length goes to zero causing more than four edges to meet at the merged cell vertex. This triggers an abrupt T1 topological transition (Figure~\ref{Topo}). The shape changes that result from the T1 can cause other edge lengths to vanish, and trigger additional T1s; the T1 cascade terminates when Plateau's Laws are fulfilled and equilibrium is restored. The instantaneous rearrangement causes a drop discontinuity in the macroscopic stress. Subsequent deformation is elastic until the occurrence of further T1 cascades, etc.

Elastic response punctuated by discrete cell-neighbor switching is a hallmark of the simple 2D Princen model of a liquid honeycomb in simple shear~\cite{Princen:83}; and also has been analyzed in 3D for the Kelvin foam~\cite{Kelvin} and Weaire-Phelan foam~\cite{WeairePhelan:94} under shear~\cite{ReineltKraynik:96,ReineltKraynik:00}. Small systems, however, are incapable of exhibiting a stress plateau in 2D~\cite{KraynikHansen:86} or in 3D, because of constraints due to spatial periodicity and flow kinematics. Simulations of disordered 2D foams~\cite{WeaireKermode:84,WeaireFortes:94,HohlerCohenAddad:05} do exhibit a stress plateau, which is a piecewise continuous function that fluctuates about its average value. The stress for disordered 2D foams increases during small elastic deformations and is partially relieved by drop discontinuities caused by T1 cascades. Large cascades can result in strain localization, also called shear banding~\cite{Goddard:03}. 

Experimental studies have addressed shear banding in quasi-2D foams, which consist of bubbles confined to a single layer in various ways~\cite{Debregas:01,Wang:06,Katgert:08,Katgert:09}. The shear bands appear to be caused by stress inhomogeneities or by viscous drag from confining glass plates~\cite{SchallvanHecke:10}. Shear bands have also been found in simulations of 2D foams in straight channels by analyzing spatial and temporal correlations of topological transitions~\cite{CoxWeaireGlazier:04,Kabla:07:1,Kabla:07:2,WynDaviesCox:08}; however, simulations of fully periodic (unbounded) 2D foams under homogeneous shear have not been reported. 

Diffusing wave spectroscopy has been used to detect intermittent shear-induced bubble rearrangements during steady flow of shaving foam~\cite{GopalDurian:99,GittingsDurian:08,LeMerrer:12}. Topological transitions have also been directly observed~\cite{Rouyer:03} in a soap froth during steady shearing flow. Experiments have yet to detect shear banding in dry 3D foams~\cite{Rodts:05,Ovarlez:10,Ovarlez:12}. 

A method for simulating the equilibrium microstructure of random soap froth with controlled cell-size distributions~\cite{Kraynik:03,Kraynik:04,Kraynik:05} employs molecular dynamics to generate dense packings of rigid spheres, which are used to produce Laguerre (weighted-Voronoi) tessellations. The tessellations are initial conditions for the Surface Evolver~\cite{Brakke:92}, which minimizes surface area given cell volume constraints. Convergence to a local surface area minimum requires numerous T1s that are triggered by cell edges going to zero length. Annealing via large-deformation, tension-compression cycles further reduces surface area and the number of short cell edges. Finally, if isotropic stress is desired, a slight distortion (recoil) of the cubic unit cell is required. The topological statistics of the resulting random monodisperse foams are in excellent agreement with Matzke's seminal experiments~\cite{Matzke}. 

Quasistatic simple shearing flow is implemented in the Surface Evolver by repeatedly subjecting a periodic foam to small strain steps (e.g., $\Delta \gamma$ = 0.005), which involve affine deformation followed by equilibration to a state that once again satisfies Plateau's laws. If any cell edge length is less than some cutoff, that edge is eliminated and topological transitions are implemented until equilibrium is achieved. The cutoff is typically between 1\% and 5\% of the average edge length in the undeformed structure (where the total shear strain $\gamma$ = 0).

\begin{figure}
\includegraphics[width = 0.99\linewidth]{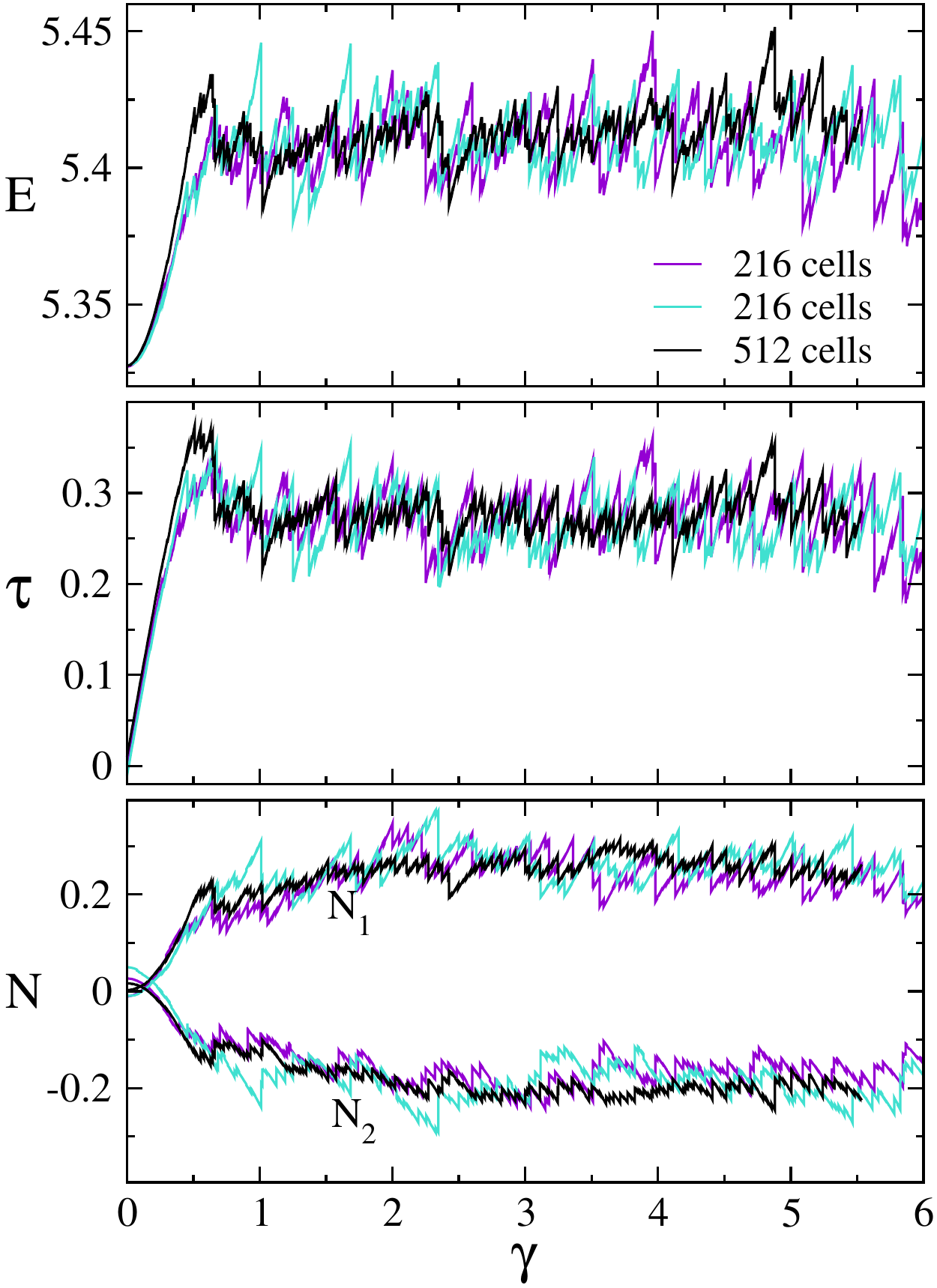}
\caption{The foam energy E (top), shear stress $\tau$ (middle), and normal stress differences $N_1$ and $N_2$ (bottom), normalized by surface tension and volume $\sigma V^{-\frac{1}{3}}$, plotted as a function of the total shear strain $\gamma$. Each plot contains two simulations with 216 ($6^3$) cells and one with 512 ($8^3$) cells. For these simulations, the initial unit cell is cubic, and therefore, $\tau$, $N_1$ and $N_2$ are non-zero at $\gamma=0$.}
\label{Rheology}
\end{figure}

The non-isotropic part of the effective macroscopic stress $\tau_{ij}$ of a dry soap froth~\cite{Kraynik:03} is evaluated as 
\begin{equation}
\tau_{ij} = \frac{2 \sigma}{V_F} \ \int_{S} ( \frac{1}{3} \ \delta_{ij} - n_i n_j ) \ ds \,,
\end{equation}
where $S$ is the set of all surfaces that divide the foam into cells. Here, $\sigma$ is the surface tension; $V_F$ is the volume of the foam; $\delta_{ij}$ is the Kronecker delta; $n_{i}$ is a local unit vector normal to the surface $S$; and $ds$ is the differential area element. The factor of 2 occurs because each surface (soap film) has two sides. 

The foam energy density is defined as $E = \sigma S_F/V_F$, where $S_F$ is the total surface area of the foam (counting both sides of each film). Rheological quantities of interest include the shear stress ($\tau=\tau_{12}$) and the first and second normal stress differences ($N_1=\tau_{11}-\tau_{22}$, $N_2=\tau_{22}-\tau_{33}$), where flow in the $1$-direction depends on the $2$-direction. Figure \ref{Rheology} shows the dependence of the energy and other rheological functions on the total strain $\gamma$.  

\begin{figure*}
\includegraphics[width = 0.28\textwidth]{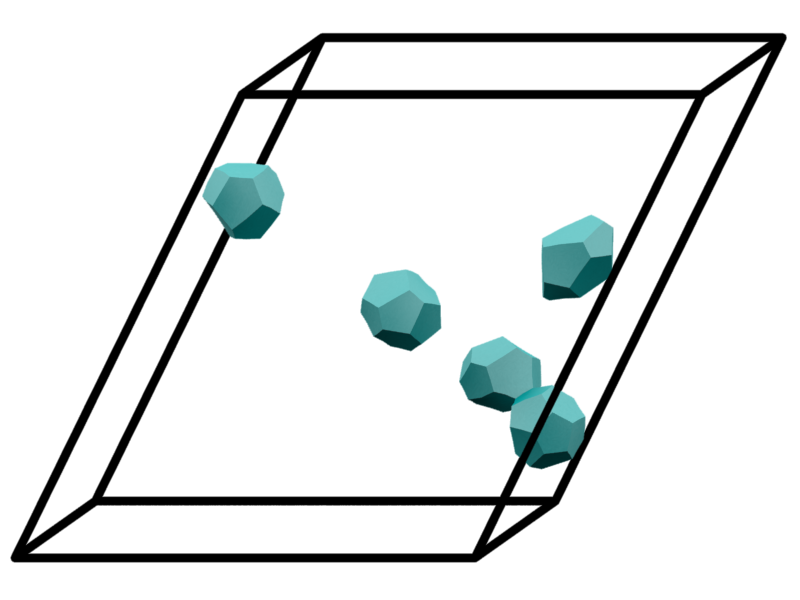}
\includegraphics[width = 0.28\textwidth]{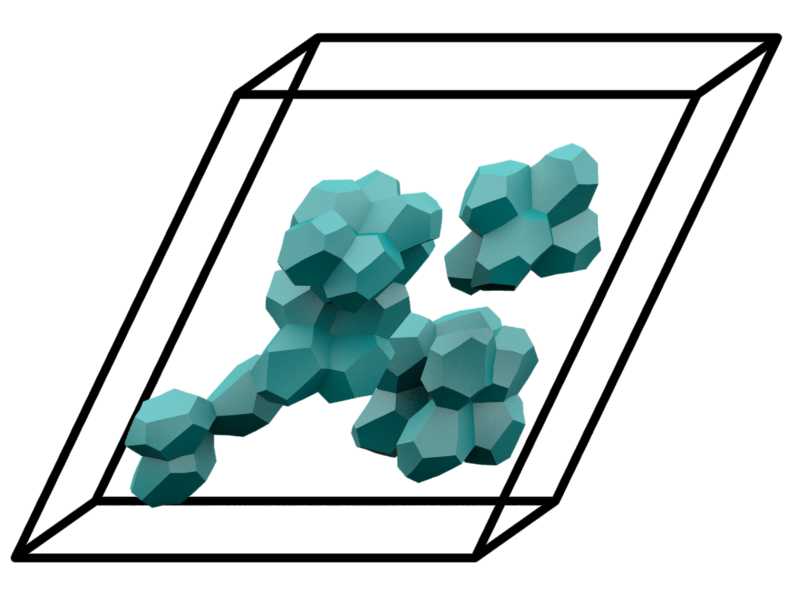}
\includegraphics[width = 0.28\textwidth]{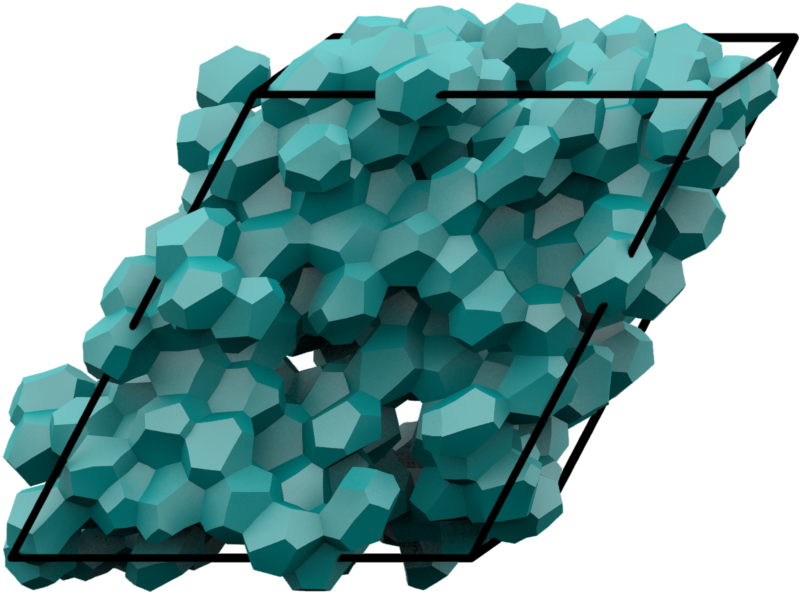}
\caption{(Left) A typical stress release domain for a purely elastic strain step consists of a few (normally between 3 and 10) disconnected cells. (Center) A typical plastic zone is composed of all cells that change topology across a topological transition. (Right) A typical stress release domain across a topological transition can contain hundreds of cells, and spans the entire foam. The foam has 512 cells and the strain is approximately 0.5.}
\label{Domains}
\end{figure*}

The foam response is initially elastic with $\tau \sim \gamma$ and $E,N_1,N_2 \sim \gamma^2$. Eventually, all quantities are expected to reach a plateau (to within fluctuations). The strain at which the first T1 occurs is system dependent and defines the elastic limit and the onset of irreversibility. The elastic limit occurs at smaller strains and lower stress than the plateau, and there is some evidence of a shear stress overshoot, which would correspond to a static yield stress. In the stress plateau regime, the stress is a piecewise continuous function, composed of small elastic deformations followed by drop discontinuities caused by T1s. 

The normal stress differences reach a plateau more gradually than the shear stress, suggesting greater sensitivity to the foam microstructure. The ratio of $-N_2$ to $N_1$ in the initial elastic regime is slightly less than $1$, comparable to the $\frac{6}{7}$ predicted in~\cite{Hohler:04}. This ratio is peculiar to foams: the ratio is generally between 0.1 and 0.3 for polymer melts and solutions~\cite{Bird:87,Larson:88}.

A measure of shape anisotropy for a given cell can be derived from the interface tensor~\cite{Kraynik:03,Evans:12}
\begin{equation}
 q_{ij} = V^{-2/3} \int_{S} \left (\frac{1}{3} \ \delta_{ij} - n_i n_j \right ) ds \,,
\end{equation}
where $S$ refers to the surface of the cell and $V$ its volume. The magnitude (second invariant) of $q_{ij}$, a dimensionless scalar $Q$ defined by
\begin{equation}
 Q = \bigl( \; \small{\sum_{i} \sum_{j}} \frac{1}{2} \ q_{ij} \ q_{ij} \bigl)^{1/2} \,,
\end{equation}
provides a simple gauge of shape anisotropy for an individual cell. Figure \ref{Colour_by_Q} shows the variation of each cell's $Q$ in a foam at two different strains; $Q$ typically ranges between 0 and 1. Note that large Q values are typical in undeformed foams where the macroscopic stress is essentially zero, which indicates the individual cells are quite deformed. When the macroscopic stress is large, the cells are deformed and aligned.

We consider the cell anisotropy changes during shear deformation of a disordered monodisperse foam with 512 cells. The increment $\Delta Q$ for each cell is computed at elastic strain steps and across T1 cascades. Cells that become more isotropic ($\Delta Q < 0$) experience \textit{stress release}. For elastic strain steps, the \textit{release domain}, the set of all cells with $\Delta Q < 0$, typically contains between 3 and 10 cells (Figure~\ref{Domains}, left). The geometric transformation at a T1 is more complicated. The set of cells that change topology across a T1, the \textit{plastic zone}, can contain between 4 and 150 cells (Figure~\ref{Domains}, center). The discrete changes in cell shape during T1 cascades significantly redistribute the cell level stress. The stress release domain across a T1 cascade contains considerably more cells (typically around 200), spanning the foam in all directions (Figure~\ref{Domains}, right). The complex morphology of this set of cells is a percolating network-like structure. 

Network-like geometry of a surface can be quantified by the Euler characteristic $\chi$~\cite{Mecke:98,Arns:01}, a topological invariant of a surface related to the number of holes and handles in the surface. It is equivalent to $2-2g$, where $g$ is the genus of the surface. For example, $\chi=2$ for a sphere and $\chi=0$ for a torus. The invariant $\chi$ of a surface $S$ can be computed as
\begin{equation}
 \chi = \frac{1}{2\pi} \int_{S} K ds \,,
\end{equation}
where $K$ is the Gaussian curvature. For structures resulting from random spatial processes, large negative values of $\chi$ are indicative of a network-like structure~\cite{Mickel:08}. 

\begin{figure}
\includegraphics[width = 0.98\columnwidth]{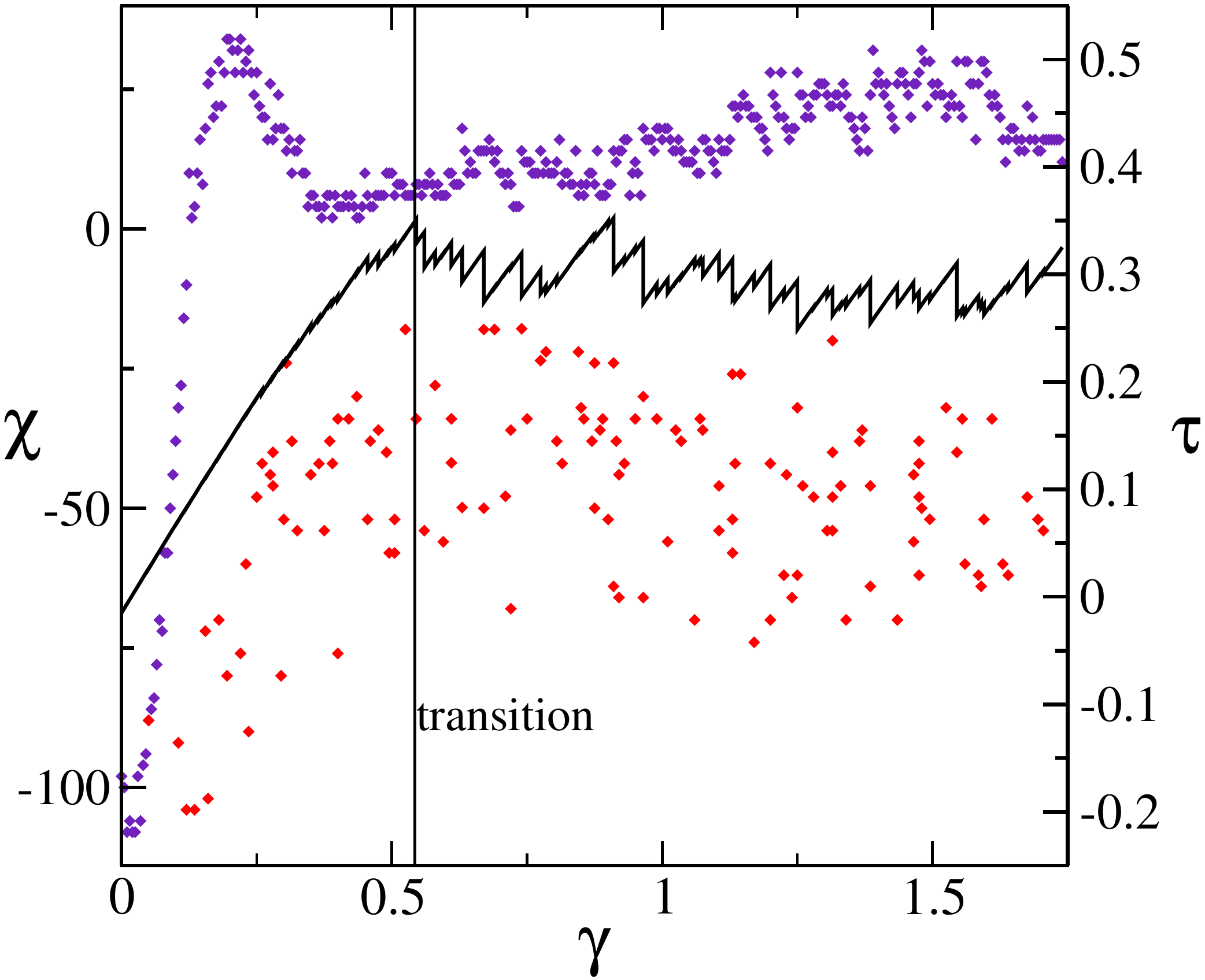}
\caption{(Color online) The distribution of $\chi$ of the stress release domains for a foam with 512 cells sheared to a strain of 1.75. The black line shows the shear stress ($\tau(\gamma)$), which reaches a plateau at $\gamma \approx 0.5$. The purple and red points show the Euler characteristic $\chi$ of the stress release domains at each elastic strain step and topological transition respectively. $\chi$  forms two distinct groups: in the stress plateau, $\chi>0$ at elastic deformations, and $\chi<0$ at T1 cascades.}
\label{Euler_plot}
\end{figure}

Figure~\ref{Euler_plot} shows the distribution of $\chi$ of the stress release domains for a shear simulation with 512 cells. The values of $\chi$ form two distinct groups: those at an elastic strain step (purple) and those across a topological transition (red). The $\chi$ values at elastic strain steps follow a relatively smooth curve that peaks prior to the transition to the stress plateau. In an isotropic foam, the principle eigenvectors of the cells are randomly distributed. When the foam is sheared, the cells begin to align along the principle eigenvector of the macroscopic stress. The observed peak in $\chi$ signifies the progression of the cells towards alignment, and is a signature of large elastic deformation and reorientation of the microstructure.  

\begin{figure}
\includegraphics[height = 0.48\columnwidth]{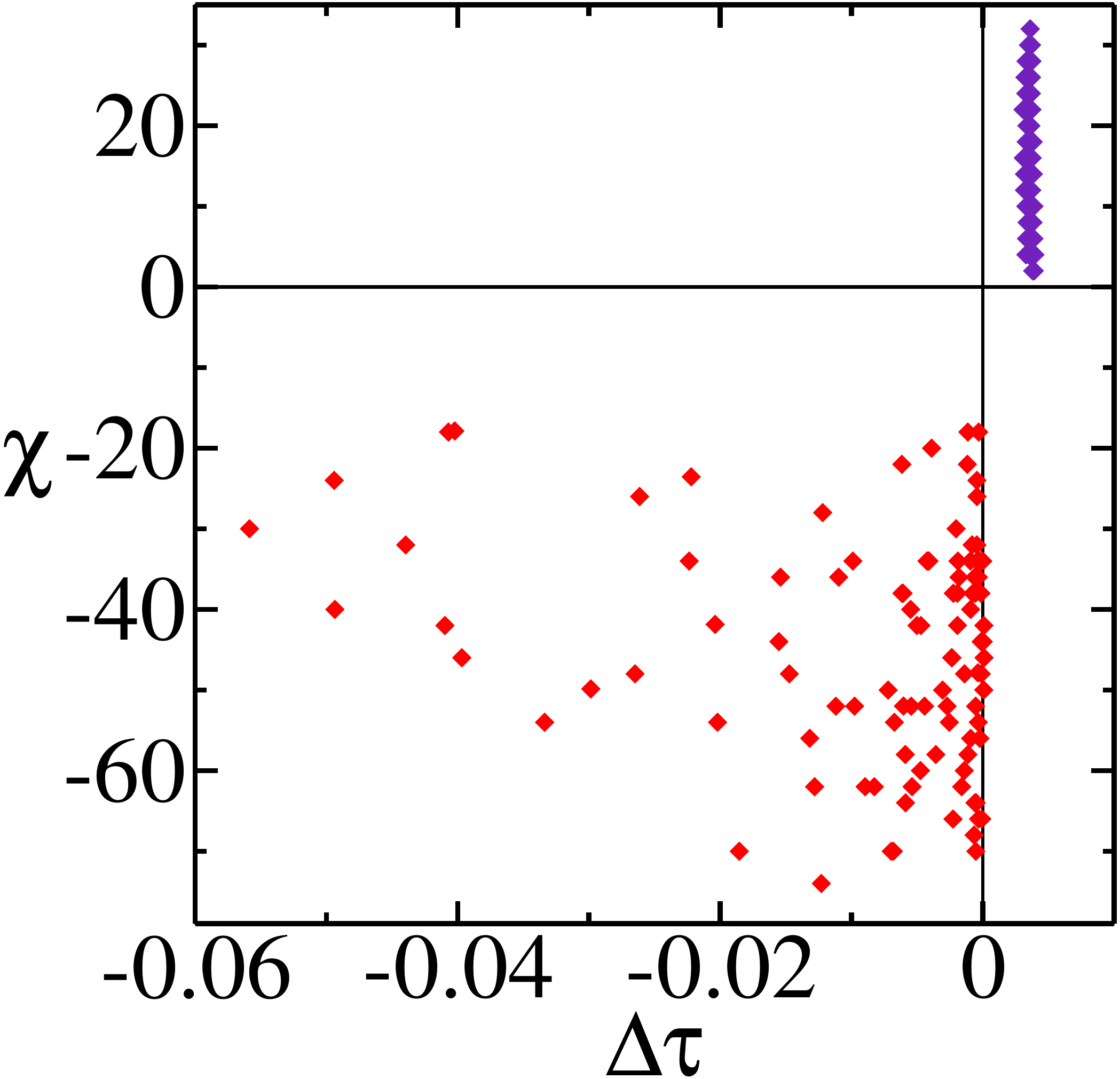}
\includegraphics[height = 0.48\columnwidth]{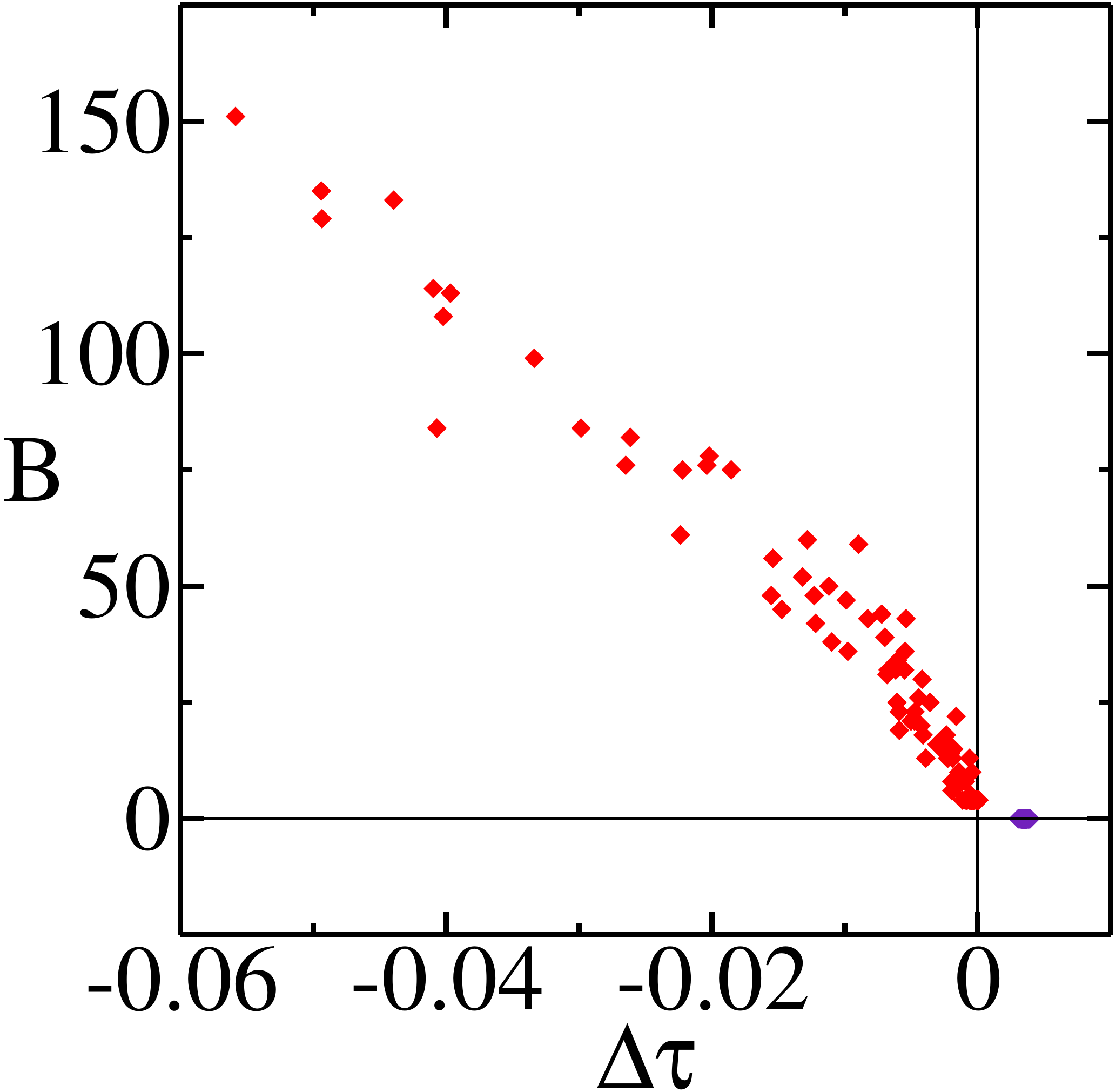}
\caption{(Color online)(Left) $\chi$ versus the change in stress $\Delta\tau$ at each elastic strain step (purple, both positive) and T1 cascade (red, both negative) for $\gamma \geq 0.5$. (Right) $\Delta\tau$ versus the size of the plastic zone $B$, approximated here by cells that change number of vertices, edges or faces, for $\gamma \geq 0.5$. Drops in stress coincide with T1 cascades, and hence negative $\chi$ indicates T1 cascades.}
\label{Stress_drop_plots}
\end{figure}

\begin{figure}
\includegraphics[width = 0.42\columnwidth]{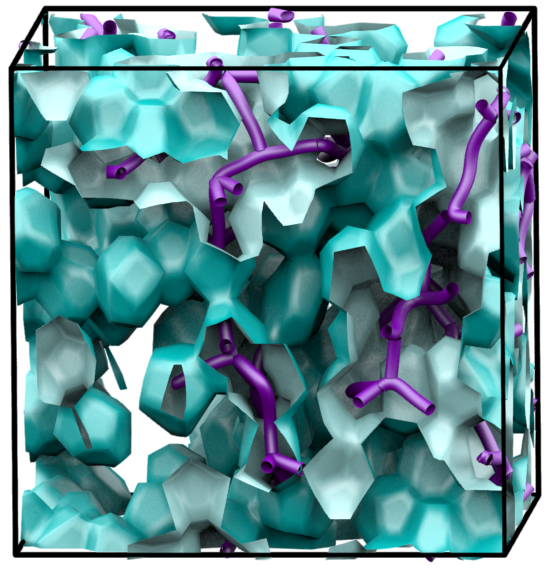}
\includegraphics[width = 0.42\columnwidth]{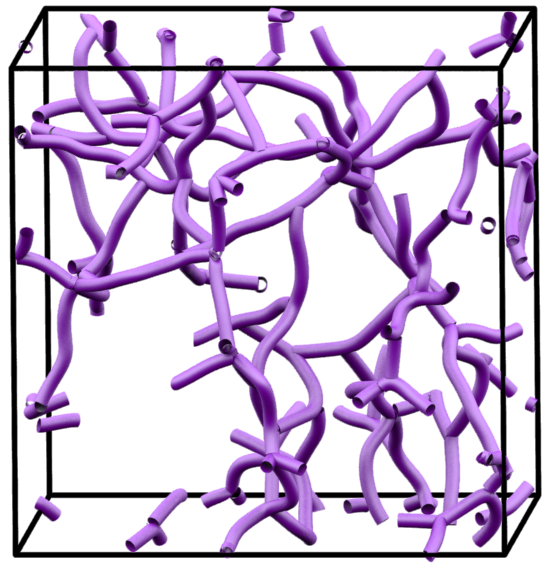}
\caption{(Color online)(Left) The surface which bounds the network-like stress release domain of across a T1 cascade is shown in green, and the medial axis representation of this domain is shown in purple. Here $\chi=-70$. (Right) The medial axis representation of the surface clearly reveals its network-like nature. The medial axis was computed by distance ordering homotopic thinning of a voxelized image~\cite{Sheppard:05,Mickel:08}.}
\label{Network_domain}
\end{figure}

Once the shear stress reaches a plateau ($\gamma \gtrsim 0.5$), $\chi$ is negative at, and only at, T1 cascades. The distinction between elastic deformations and T1 cascades is again highlighted in Figure~\ref{Stress_drop_plots}, where each drop discontinuity in shear stress coincides with negative $\chi$ and a T1 cascade. These correlations confirm the complex network-like structure of the stress release domains across T1 cascades, such as the example shown in Figure~\ref{Network_domain}. The combination of these results is a striking demonstration of the highly non-local effects of irreversible T1 cascades. The observation that a T1 cascade releases stress along a connected network spanning the foam is reminiscent of force chains and force networks in granular media~\cite{Liu:95,Cates:98}.

Since simulations of significantly larger systems are not currently feasible with the \textit{Surface Evolver}, the question of a long-term decay of the stress release domain cannot be addressed here. However, it appears likely that for cell numbers typical for experimental foam studies, the stress release reverberates through the entire system.

In conclusion, we have demonstrated that Surface Evolver simulations of quasistatic shearing flow can provide detailed insight into the processes that lead to elastic-plastic behavior of random soap froth. Our key observation is that the stress release associated with plastic flow events reverberates macroscopically throughout the foam sample, along a percolating network-like stress release domain. This highlights the complex rheology of foam, where discrete local events cause global geometric changes along with global redistribution of stress.

\begin{acknowledgments}

We acknowledge funding by the DFG through the research group `Geometry and Physics of Spatial Random Systems' (SCHR1148/3-1) and, for travel support of AMK, through the `Engineering of Advanced Materials' cluster of excellence. MEE acknowledges funding by the Humboldt foundation. We thank Kenneth Brakke for his \textit{Surface Evolver}, Adrian Sheppard for the medial axis representation in Figure~\ref{Network_domain}(b) and Sylvie {Cohen-Addad} and Reinhard H\"{o}hler for comments.

\end{acknowledgments}



%


\begin{thebibliography}{45}%
\makeatletter
\providecommand \@ifxundefined [1]{%
 \@ifx{#1\undefined}
}%
\providecommand \@ifnum [1]{%
 \ifnum #1\expandafter \@firstoftwo
 \else \expandafter \@secondoftwo
 \fi
}%
\providecommand \@ifx [1]{%
 \ifx #1\expandafter \@firstoftwo
 \else \expandafter \@secondoftwo
 \fi
}%
\providecommand \natexlab [1]{#1}%
\providecommand \enquote  [1]{``#1''}%
\providecommand \bibnamefont  [1]{#1}%
\providecommand \bibfnamefont [1]{#1}%
\providecommand \citenamefont [1]{#1}%
\providecommand \href@noop [0]{\@secondoftwo}%
\providecommand \href [0]{\begingroup \@sanitize@url \@href}%
\providecommand \@href[1]{\@@startlink{#1}\@@href}%
\providecommand \@@href[1]{\endgroup#1\@@endlink}%
\providecommand \@sanitize@url [0]{\catcode `\\12\catcode `\$12\catcode
  `\&12\catcode `\#12\catcode `\^12\catcode `\_12\catcode `\%12\relax}%
\providecommand \@@startlink[1]{}%
\providecommand \@@endlink[0]{}%
\providecommand \url  [0]{\begingroup\@sanitize@url \@url }%
\providecommand \@url [1]{\endgroup\@href {#1}{\urlprefix }}%
\providecommand \urlprefix  [0]{URL }%
\providecommand \Eprint [0]{\href }%
\providecommand \doibase [0]{http://dx.doi.org/}%
\providecommand \selectlanguage [0]{\@gobble}%
\providecommand \bibinfo  [0]{\@secondoftwo}%
\providecommand \bibfield  [0]{\@secondoftwo}%
\providecommand \translation [1]{[#1]}%
\providecommand \BibitemOpen [0]{}%
\providecommand \bibitemStop [0]{}%
\providecommand \bibitemNoStop [0]{.\EOS\space}%
\providecommand \EOS [0]{\spacefactor3000\relax}%
\providecommand \BibitemShut  [1]{\csname bibitem#1\endcsname}%
\let\auto@bib@innerbib\@empty
\bibitem [{\citenamefont {Kraynik}(1988)}]{Kraynik:88}%
  \BibitemOpen
  \bibfield  {author} {\bibinfo {author} {\bibfnamefont {A.~M.}\ \bibnamefont
  {Kraynik}},\ }\href@noop {} {\bibfield  {journal} {\bibinfo  {journal} {Annu.
  Rev. Fluid Mech.}\ }\textbf {\bibinfo {volume} {20}},\ \bibinfo {pages} {325}
  (\bibinfo {year} {1988})}\BibitemShut {NoStop}%
\bibitem [{\citenamefont {H\"ohler}\ and\ \citenamefont
  {Cohen-Addad}(2005)}]{HohlerCohenAddad:05}%
  \BibitemOpen
  \bibfield  {author} {\bibinfo {author} {\bibfnamefont {R.}~\bibnamefont
  {H\"ohler}}\ and\ \bibinfo {author} {\bibfnamefont {S.}~\bibnamefont
  {Cohen-Addad}},\ }\href@noop {} {\bibfield  {journal} {\bibinfo  {journal}
  {J. Phys.: Condens. Matter}\ }\textbf {\bibinfo {volume} {17}},\ \bibinfo
  {pages} {R1041} (\bibinfo {year} {2005})}\BibitemShut {NoStop}%
\bibitem [{\citenamefont {Princen}(1985)}]{Princen:85}%
  \BibitemOpen
  \bibfield  {author} {\bibinfo {author} {\bibfnamefont {H.~M.}\ \bibnamefont
  {Princen}},\ }\href@noop {} {\bibfield  {journal} {\bibinfo  {journal} {J.
  Colloid Interf. Sci.}\ }\textbf {\bibinfo {volume} {105}},\ \bibinfo {pages}
  {150} (\bibinfo {year} {1985})}\BibitemShut {NoStop}%
\bibitem [{\citenamefont {Princen}\ and\ \citenamefont
  {Kiss}(1986)}]{Princen:86}%
  \BibitemOpen
  \bibfield  {author} {\bibinfo {author} {\bibfnamefont {H.~M.}\ \bibnamefont
  {Princen}}\ and\ \bibinfo {author} {\bibfnamefont {A.~D.}\ \bibnamefont
  {Kiss}},\ }\href@noop {} {\bibfield  {journal} {\bibinfo  {journal} {J.
  Colloid Interface Sci.}\ }\textbf {\bibinfo {volume} {112}},\ \bibinfo
  {pages} {427} (\bibinfo {year} {1986})}\BibitemShut {NoStop}%
\bibitem [{\citenamefont {Plateau}(1873)}]{Plateau}%
  \BibitemOpen
  \bibfield  {author} {\bibinfo {author} {\bibfnamefont {J.~A.~F.}\
  \bibnamefont {Plateau}},\ }\href@noop {} {\emph {\bibinfo {title} {Statique
  Experimentale et Theorique des Liquides}}}\ (\bibinfo  {publisher}
  {Gauthier-Villiard, Paris},\ \bibinfo {year} {1873})\BibitemShut {NoStop}%
\bibitem [{\citenamefont {Princen}(1983)}]{Princen:83}%
  \BibitemOpen
  \bibfield  {author} {\bibinfo {author} {\bibfnamefont {H.~M.}\ \bibnamefont
  {Princen}},\ }\href@noop {} {\bibfield  {journal} {\bibinfo  {journal} {J.
  Colloid Interf. Sci.}\ }\textbf {\bibinfo {volume} {91}},\ \bibinfo {pages}
  {160} (\bibinfo {year} {1983})}\BibitemShut {NoStop}%
\bibitem [{\citenamefont {Kelvin}(1887)}]{Kelvin}%
  \BibitemOpen
  \bibfield  {author} {\bibinfo {author} {\bibfnamefont {L.~W.~T.}\
  \bibnamefont {Kelvin}},\ }\href@noop {} {\bibfield  {journal} {\bibinfo
  {journal} {Phil. Mag.}\ }\textbf {\bibinfo {volume} {24}},\ \bibinfo {pages}
  {503} (\bibinfo {year} {1887})}\BibitemShut {NoStop}%
\bibitem [{\citenamefont {Weaire}\ and\ \citenamefont
  {Phelan}(1994)}]{WeairePhelan:94}%
  \BibitemOpen
  \bibfield  {author} {\bibinfo {author} {\bibfnamefont {D.}~\bibnamefont
  {Weaire}}\ and\ \bibinfo {author} {\bibfnamefont {R.}~\bibnamefont
  {Phelan}},\ }\href@noop {} {\bibfield  {journal} {\bibinfo  {journal} {Phil.
  Mag. Lett.}\ }\textbf {\bibinfo {volume} {69}},\ \bibinfo {pages} {107}
  (\bibinfo {year} {1994})}\BibitemShut {NoStop}%
\bibitem [{\citenamefont {Reinelt}\ and\ \citenamefont
  {Kraynik}(1996)}]{ReineltKraynik:96}%
  \BibitemOpen
  \bibfield  {author} {\bibinfo {author} {\bibfnamefont {D.~A.}\ \bibnamefont
  {Reinelt}}\ and\ \bibinfo {author} {\bibfnamefont {A.~M.}\ \bibnamefont
  {Kraynik}},\ }\href@noop {} {\bibfield  {journal} {\bibinfo  {journal} {J.
  Fluid Mech.}\ }\textbf {\bibinfo {volume} {311}},\ \bibinfo {pages} {327}
  (\bibinfo {year} {1996})}\BibitemShut {NoStop}%
\bibitem [{\citenamefont {Reinelt}\ and\ \citenamefont
  {Kraynik}(2000)}]{ReineltKraynik:00}%
  \BibitemOpen
  \bibfield  {author} {\bibinfo {author} {\bibfnamefont {D.~A.}\ \bibnamefont
  {Reinelt}}\ and\ \bibinfo {author} {\bibfnamefont {A.~M.}\ \bibnamefont
  {Kraynik}},\ }\href@noop {} {\bibfield  {journal} {\bibinfo  {journal} {J.
  Rheol.}\ }\textbf {\bibinfo {volume} {44}},\ \bibinfo {pages} {453} (\bibinfo
  {year} {2000})}\BibitemShut {NoStop}%
\bibitem [{\citenamefont {Kraynik}\ and\ \citenamefont
  {Hansen}(1986)}]{KraynikHansen:86}%
  \BibitemOpen
  \bibfield  {author} {\bibinfo {author} {\bibfnamefont {A.~M.}\ \bibnamefont
  {Kraynik}}\ and\ \bibinfo {author} {\bibfnamefont {M.~G.}\ \bibnamefont
  {Hansen}},\ }\href@noop {} {\bibfield  {journal} {\bibinfo  {journal} {J.
  Rheol.}\ }\textbf {\bibinfo {volume} {30}},\ \bibinfo {pages} {409} (\bibinfo
  {year} {1986})}\BibitemShut {NoStop}%
\bibitem [{\citenamefont {Weaire}\ and\ \citenamefont
  {Kermode}(1984)}]{WeaireKermode:84}%
  \BibitemOpen
  \bibfield  {author} {\bibinfo {author} {\bibfnamefont {D.}~\bibnamefont
  {Weaire}}\ and\ \bibinfo {author} {\bibfnamefont {J.~P.}\ \bibnamefont
  {Kermode}},\ }\href@noop {} {\bibfield  {journal} {\bibinfo  {journal} {Phil.
  Mag. B}\ }\textbf {\bibinfo {volume} {50}},\ \bibinfo {pages} {379} (\bibinfo
  {year} {1984})}\BibitemShut {NoStop}%
\bibitem [{\citenamefont {Weaire}\ and\ \citenamefont
  {Fortes}(1994)}]{WeaireFortes:94}%
  \BibitemOpen
  \bibfield  {author} {\bibinfo {author} {\bibfnamefont {D.}~\bibnamefont
  {Weaire}}\ and\ \bibinfo {author} {\bibfnamefont {M.~A.}\ \bibnamefont
  {Fortes}},\ }\href@noop {} {\bibfield  {journal} {\bibinfo  {journal} {Adv.
  Phys.}\ }\textbf {\bibinfo {volume} {43}},\ \bibinfo {pages} {685} (\bibinfo
  {year} {1994})}\BibitemShut {NoStop}%
\bibitem [{\citenamefont {Goddard}(2003)}]{Goddard:03}%
  \BibitemOpen
  \bibfield  {author} {\bibinfo {author} {\bibfnamefont {J.~D.}\ \bibnamefont
  {Goddard}},\ }\href@noop {} {\bibfield  {journal} {\bibinfo  {journal} {Annu.
  Rev. Fluid Mech.}\ }\textbf {\bibinfo {volume} {35}},\ \bibinfo {pages} {113}
  (\bibinfo {year} {2003})}\BibitemShut {NoStop}%
\bibitem [{\citenamefont {Debregeas}\ \emph {et~al.}(2001)\citenamefont
  {Debregeas}, \citenamefont {Tabuteau},\ and\ \citenamefont
  {di~Meglio}}]{Debregas:01}%
  \BibitemOpen
  \bibfield  {author} {\bibinfo {author} {\bibfnamefont {G.}~\bibnamefont
  {Debregeas}}, \bibinfo {author} {\bibfnamefont {H.}~\bibnamefont {Tabuteau}},
  \ and\ \bibinfo {author} {\bibfnamefont {J.~M.}\ \bibnamefont {di~Meglio}},\
  }\href@noop {} {\bibfield  {journal} {\bibinfo  {journal} {Phys. Rev. Lett.}\
  }\textbf {\bibinfo {volume} {87}},\ \bibinfo {pages} {178305} (\bibinfo
  {year} {2001})}\BibitemShut {NoStop}%
\bibitem [{\citenamefont {Wang}\ \emph {et~al.}(2006)\citenamefont {Wang},
  \citenamefont {Krishan},\ and\ \citenamefont {Dennin}}]{Wang:06}%
  \BibitemOpen
  \bibfield  {author} {\bibinfo {author} {\bibfnamefont {Y.~H.}\ \bibnamefont
  {Wang}}, \bibinfo {author} {\bibfnamefont {K.}~\bibnamefont {Krishan}}, \
  and\ \bibinfo {author} {\bibfnamefont {M.}~\bibnamefont {Dennin}},\
  }\href@noop {} {\bibfield  {journal} {\bibinfo  {journal} {Phys. Rev. E}\
  }\textbf {\bibinfo {volume} {73}},\ \bibinfo {pages} {031401} (\bibinfo
  {year} {2006})}\BibitemShut {NoStop}%
\bibitem [{\citenamefont {Katgert}\ \emph {et~al.}(2008)\citenamefont
  {Katgert}, \citenamefont {Mobius},\ and\ \citenamefont {van
  Hecke}}]{Katgert:08}%
  \BibitemOpen
  \bibfield  {author} {\bibinfo {author} {\bibfnamefont {G.}~\bibnamefont
  {Katgert}}, \bibinfo {author} {\bibfnamefont {M.~E.}\ \bibnamefont {Mobius}},
  \ and\ \bibinfo {author} {\bibfnamefont {M.}~\bibnamefont {van Hecke}},\
  }\href@noop {} {\bibfield  {journal} {\bibinfo  {journal} {Phys. Rev. Lett.}\
  }\textbf {\bibinfo {volume} {101}},\ \bibinfo {pages} {058301} (\bibinfo
  {year} {2008})}\BibitemShut {NoStop}%
\bibitem [{\citenamefont {Katgert}\ \emph {et~al.}(2009)\citenamefont
  {Katgert}, \citenamefont {Latka}, \citenamefont {Mobius},\ and\ \citenamefont
  {van Hecke}}]{Katgert:09}%
  \BibitemOpen
  \bibfield  {author} {\bibinfo {author} {\bibfnamefont {G.}~\bibnamefont
  {Katgert}}, \bibinfo {author} {\bibfnamefont {A.}~\bibnamefont {Latka}},
  \bibinfo {author} {\bibfnamefont {M.~E.}\ \bibnamefont {Mobius}}, \ and\
  \bibinfo {author} {\bibfnamefont {M.}~\bibnamefont {van Hecke}},\ }\href@noop
  {} {\bibfield  {journal} {\bibinfo  {journal} {Phys. Rev. E}\ }\textbf
  {\bibinfo {volume} {79}},\ \bibinfo {pages} {066318} (\bibinfo {year}
  {2009})}\BibitemShut {NoStop}%
\bibitem [{\citenamefont {Schall}\ and\ \citenamefont {van
  Hecke}(2010)}]{SchallvanHecke:10}%
  \BibitemOpen
  \bibfield  {author} {\bibinfo {author} {\bibfnamefont {P.}~\bibnamefont
  {Schall}}\ and\ \bibinfo {author} {\bibfnamefont {M.}~\bibnamefont {van
  Hecke}},\ }\href@noop {} {\bibfield  {journal} {\bibinfo  {journal} {Annu.
  Rev. Fluid Mech.}\ }\textbf {\bibinfo {volume} {42}},\ \bibinfo {pages} {67}
  (\bibinfo {year} {2010})}\BibitemShut {NoStop}%
\bibitem [{\citenamefont {Cox}\ \emph {et~al.}(2004)\citenamefont {Cox},
  \citenamefont {Weaire},\ and\ \citenamefont {Glazier}}]{CoxWeaireGlazier:04}%
  \BibitemOpen
  \bibfield  {author} {\bibinfo {author} {\bibfnamefont {S.}~\bibnamefont
  {Cox}}, \bibinfo {author} {\bibfnamefont {D.}~\bibnamefont {Weaire}}, \ and\
  \bibinfo {author} {\bibfnamefont {J.~A.}\ \bibnamefont {Glazier}},\
  }\href@noop {} {\bibfield  {journal} {\bibinfo  {journal} {Rheol. Acta}\
  }\textbf {\bibinfo {volume} {43}},\ \bibinfo {pages} {442} (\bibinfo {year}
  {2004})}\BibitemShut {NoStop}%
\bibitem [{\citenamefont {Kabla}\ and\ \citenamefont
  {Debregas}(2007)}]{Kabla:07:1}%
  \BibitemOpen
  \bibfield  {author} {\bibinfo {author} {\bibfnamefont {A.}~\bibnamefont
  {Kabla}}\ and\ \bibinfo {author} {\bibfnamefont {G.}~\bibnamefont
  {Debregas}},\ }\href@noop {} {\bibfield  {journal} {\bibinfo  {journal} {J.
  Fluid Mech.}\ }\textbf {\bibinfo {volume} {587}},\ \bibinfo {pages} {22}
  (\bibinfo {year} {2007})}\BibitemShut {NoStop}%
\bibitem [{\citenamefont {Kabla}\ \emph {et~al.}(2007)\citenamefont {Kabla},
  \citenamefont {Scheibert},\ and\ \citenamefont {Debregas}}]{Kabla:07:2}%
  \BibitemOpen
  \bibfield  {author} {\bibinfo {author} {\bibfnamefont {A.}~\bibnamefont
  {Kabla}}, \bibinfo {author} {\bibfnamefont {J.}~\bibnamefont {Scheibert}}, \
  and\ \bibinfo {author} {\bibfnamefont {G.}~\bibnamefont {Debregas}},\
  }\href@noop {} {\bibfield  {journal} {\bibinfo  {journal} {J. Fluid Mech.}\
  }\textbf {\bibinfo {volume} {587}},\ \bibinfo {pages} {45} (\bibinfo {year}
  {2007})}\BibitemShut {NoStop}%
\bibitem [{\citenamefont {Wyn}\ \emph {et~al.}(2008)\citenamefont {Wyn},
  \citenamefont {Davies},\ and\ \citenamefont {Cox}}]{WynDaviesCox:08}%
  \BibitemOpen
  \bibfield  {author} {\bibinfo {author} {\bibfnamefont {A.}~\bibnamefont
  {Wyn}}, \bibinfo {author} {\bibfnamefont {I.~T.}\ \bibnamefont {Davies}}, \
  and\ \bibinfo {author} {\bibfnamefont {S.~J.}\ \bibnamefont {Cox}},\
  }\href@noop {} {\bibfield  {journal} {\bibinfo  {journal} {Eur. Phys. J. E}\
  }\textbf {\bibinfo {volume} {26}},\ \bibinfo {pages} {81} (\bibinfo {year}
  {2008})}\BibitemShut {NoStop}%
\bibitem [{\citenamefont {Gopal}\ and\ \citenamefont
  {Durian}(1999)}]{GopalDurian:99}%
  \BibitemOpen
  \bibfield  {author} {\bibinfo {author} {\bibfnamefont {A.~D.}\ \bibnamefont
  {Gopal}}\ and\ \bibinfo {author} {\bibfnamefont {D.~J.}\ \bibnamefont
  {Durian}},\ }\href@noop {} {\bibfield  {journal} {\bibinfo  {journal} {J.
  Colloid Interf. Sci.}\ }\textbf {\bibinfo {volume} {213}},\ \bibinfo {pages}
  {169} (\bibinfo {year} {1999})}\BibitemShut {NoStop}%
\bibitem [{\citenamefont {Gittings}\ and\ \citenamefont
  {Durian}(2008)}]{GittingsDurian:08}%
  \BibitemOpen
  \bibfield  {author} {\bibinfo {author} {\bibfnamefont {A.~S.}\ \bibnamefont
  {Gittings}}\ and\ \bibinfo {author} {\bibfnamefont {D.~J.}\ \bibnamefont
  {Durian}},\ }\href@noop {} {\bibfield  {journal} {\bibinfo  {journal} {Phys.
  Rev. E}\ }\textbf {\bibinfo {volume} {78}},\ \bibinfo {pages} {066313}
  (\bibinfo {year} {2008})}\BibitemShut {NoStop}%
\bibitem [{\citenamefont {{Le Merrer}}\ \emph {et~al.}(2012)\citenamefont {{Le
  Merrer}}, \citenamefont {Cohen-Addad},\ and\ \citenamefont
  {H\"ohler}}]{LeMerrer:12}%
  \BibitemOpen
  \bibfield  {author} {\bibinfo {author} {\bibfnamefont {M.}~\bibnamefont {{Le
  Merrer}}}, \bibinfo {author} {\bibfnamefont {S.}~\bibnamefont {Cohen-Addad}},
  \ and\ \bibinfo {author} {\bibfnamefont {R.}~\bibnamefont {H\"ohler}},\
  }\href@noop {} {\bibfield  {journal} {\bibinfo  {journal} {Phys. Rev. Lett.}\
  }\textbf {\bibinfo {volume} {108}},\ \bibinfo {pages} {188301} (\bibinfo
  {year} {2012})}\BibitemShut {NoStop}%
\bibitem [{\citenamefont {Rouyer}\ \emph {et~al.}(2003)\citenamefont {Rouyer},
  \citenamefont {Cohen-Addad}, \citenamefont {Vignes-Adler},\ and\
  \citenamefont {H\"ohler}}]{Rouyer:03}%
  \BibitemOpen
  \bibfield  {author} {\bibinfo {author} {\bibfnamefont {F.}~\bibnamefont
  {Rouyer}}, \bibinfo {author} {\bibfnamefont {S.}~\bibnamefont {Cohen-Addad}},
  \bibinfo {author} {\bibfnamefont {M.}~\bibnamefont {Vignes-Adler}}, \ and\
  \bibinfo {author} {\bibfnamefont {R.}~\bibnamefont {H\"ohler}},\ }\href@noop
  {} {\bibfield  {journal} {\bibinfo  {journal} {Phys. Rev. E}\ }\textbf
  {\bibinfo {volume} {67}},\ \bibinfo {pages} {021405} (\bibinfo {year}
  {2003})}\BibitemShut {NoStop}%
\bibitem [{\citenamefont {Rodts}\ \emph {et~al.}(2005)\citenamefont {Rodts},
  \citenamefont {Baudez},\ and\ \citenamefont {Coussot}}]{Rodts:05}%
  \BibitemOpen
  \bibfield  {author} {\bibinfo {author} {\bibfnamefont {S.}~\bibnamefont
  {Rodts}}, \bibinfo {author} {\bibfnamefont {J.~C.}\ \bibnamefont {Baudez}}, \
  and\ \bibinfo {author} {\bibfnamefont {P.}~\bibnamefont {Coussot}},\
  }\href@noop {} {\bibfield  {journal} {\bibinfo  {journal} {Europhys. Lett.}\
  }\textbf {\bibinfo {volume} {69}},\ \bibinfo {pages} {636} (\bibinfo {year}
  {2005})}\BibitemShut {NoStop}%
\bibitem [{\citenamefont {Ovarlez}\ \emph {et~al.}(2010)\citenamefont
  {Ovarlez}, \citenamefont {Krishan},\ and\ \citenamefont
  {Cohen-Addad}}]{Ovarlez:10}%
  \BibitemOpen
  \bibfield  {author} {\bibinfo {author} {\bibfnamefont {G.}~\bibnamefont
  {Ovarlez}}, \bibinfo {author} {\bibfnamefont {K.}~\bibnamefont {Krishan}}, \
  and\ \bibinfo {author} {\bibfnamefont {S.}~\bibnamefont {Cohen-Addad}},\
  }\href@noop {} {\bibfield  {journal} {\bibinfo  {journal} {Europhys. Lett.}\
  }\textbf {\bibinfo {volume} {91}},\ \bibinfo {pages} {68005} (\bibinfo {year}
  {2010})}\BibitemShut {NoStop}%
\bibitem [{\citenamefont {Ovarlez}\ \emph {et~al.}(2013)\citenamefont
  {Ovarlez}, \citenamefont {S}, \citenamefont {Krishan}, \citenamefont
  {Goyon},\ and\ \citenamefont {Coussot}}]{Ovarlez:12}%
  \BibitemOpen
  \bibfield  {author} {\bibinfo {author} {\bibfnamefont {G.}~\bibnamefont
  {Ovarlez}}, \bibinfo {author} {\bibfnamefont {C.-A.}\ \bibnamefont {S}},
  \bibinfo {author} {\bibfnamefont {K.}~\bibnamefont {Krishan}}, \bibinfo
  {author} {\bibfnamefont {J.}~\bibnamefont {Goyon}}, \ and\ \bibinfo {author}
  {\bibfnamefont {P.}~\bibnamefont {Coussot}},\ }\href
  {http://dx.doi.org/10.1016/j.jnnfm.2012.06.009} {\bibfield  {journal}
  {\bibinfo  {journal} {J. Non-Newton. Fluid}\ }\textbf {\bibinfo {volume}
  {193}},\ \bibinfo {pages} {68} (\bibinfo {year} {2013})}\BibitemShut
  {NoStop}%
\bibitem [{\citenamefont {Kraynik}\ \emph {et~al.}(2003)\citenamefont
  {Kraynik}, \citenamefont {Reinelt},\ and\ \citenamefont {{van
  Swol}}}]{Kraynik:03}%
  \BibitemOpen
  \bibfield  {author} {\bibinfo {author} {\bibfnamefont {A.~M.}\ \bibnamefont
  {Kraynik}}, \bibinfo {author} {\bibfnamefont {D.~A.}\ \bibnamefont
  {Reinelt}}, \ and\ \bibinfo {author} {\bibfnamefont {F.}~\bibnamefont {{van
  Swol}}},\ }\href {\doibase 10.1103/PhysRevE.67.031403} {\bibfield  {journal}
  {\bibinfo  {journal} {Phys. Rev. E}\ }\textbf {\bibinfo {volume} {67}},\
  \bibinfo {pages} {031403} (\bibinfo {year} {2003})}\BibitemShut {NoStop}%
\bibitem [{\citenamefont {Kraynik}\ \emph {et~al.}(2004)\citenamefont
  {Kraynik}, \citenamefont {Reinelt},\ and\ \citenamefont {F.~van
  Swol}}]{Kraynik:04}%
  \BibitemOpen
  \bibfield  {author} {\bibinfo {author} {\bibfnamefont {A.~M.}\ \bibnamefont
  {Kraynik}}, \bibinfo {author} {\bibfnamefont {D.~A.}\ \bibnamefont
  {Reinelt}}, \ and\ \bibinfo {author} {\bibfnamefont {F.}~\bibnamefont {F.~van
  Swol}},\ }\href@noop {} {\bibfield  {journal} {\bibinfo  {journal} {Phys.
  Rev. Lett.}\ }\textbf {\bibinfo {volume} {93}},\ \bibinfo {pages} {208301}
  (\bibinfo {year} {2004})}\BibitemShut {NoStop}%
\bibitem [{\citenamefont {Kraynik}\ \emph {et~al.}(2005)\citenamefont
  {Kraynik}, \citenamefont {Reinelt},\ and\ \citenamefont {F.~van
  Swol}}]{Kraynik:05}%
  \BibitemOpen
  \bibfield  {author} {\bibinfo {author} {\bibfnamefont {A.~M.}\ \bibnamefont
  {Kraynik}}, \bibinfo {author} {\bibfnamefont {D.~A.}\ \bibnamefont
  {Reinelt}}, \ and\ \bibinfo {author} {\bibfnamefont {F.}~\bibnamefont {F.~van
  Swol}},\ }\href@noop {} {\bibfield  {journal} {\bibinfo  {journal} {Colloid
  Surface A}\ }\textbf {\bibinfo {volume} {263}},\ \bibinfo {pages} {11}
  (\bibinfo {year} {2005})}\BibitemShut {NoStop}%
\bibitem [{\citenamefont {Brakke}(1992)}]{Brakke:92}%
  \BibitemOpen
  \bibfield  {author} {\bibinfo {author} {\bibfnamefont {K.}~\bibnamefont
  {Brakke}},\ }\href@noop {} {\bibfield  {journal} {\bibinfo  {journal} {Exp.
  Math.}\ }\textbf {\bibinfo {volume} {1}},\ \bibinfo {pages} {141} (\bibinfo
  {year} {1992})}\BibitemShut {NoStop}%
\bibitem [{\citenamefont {Matzke}(1946)}]{Matzke}%
  \BibitemOpen
  \bibfield  {author} {\bibinfo {author} {\bibfnamefont {E.~B.}\ \bibnamefont
  {Matzke}},\ }\href@noop {} {\bibfield  {journal} {\bibinfo  {journal} {Am. J.
  Botany}\ }\textbf {\bibinfo {volume} {33}},\ \bibinfo {pages} {58} (\bibinfo
  {year} {1946})}\BibitemShut {NoStop}%
\bibitem [{\citenamefont {H\"ohler}\ \emph {et~al.}(2004)\citenamefont
  {H\"ohler}, \citenamefont {Cohen-Addad},\ and\ \citenamefont
  {Labiausse}}]{Hohler:04}%
  \BibitemOpen
  \bibfield  {author} {\bibinfo {author} {\bibfnamefont {R.}~\bibnamefont
  {H\"ohler}}, \bibinfo {author} {\bibfnamefont {S.}~\bibnamefont
  {Cohen-Addad}}, \ and\ \bibinfo {author} {\bibfnamefont {V.}~\bibnamefont
  {Labiausse}},\ }\href@noop {} {\bibfield  {journal} {\bibinfo  {journal} {J.
  Rheol.}\ }\textbf {\bibinfo {volume} {48}},\ \bibinfo {pages} {679} (\bibinfo
  {year} {2004})}\BibitemShut {NoStop}%
\bibitem [{\citenamefont {Bird}\ \emph {et~al.}(1987)\citenamefont {Bird},
  \citenamefont {Armstrong},\ and\ \citenamefont {Hassager}}]{Bird:87}%
  \BibitemOpen
  \bibfield  {author} {\bibinfo {author} {\bibfnamefont {R.~B.}\ \bibnamefont
  {Bird}}, \bibinfo {author} {\bibfnamefont {R.~C.}\ \bibnamefont {Armstrong}},
  \ and\ \bibinfo {author} {\bibfnamefont {O.}~\bibnamefont {Hassager}},\
  }\href@noop {} {\emph {\bibinfo {title} {Dynamics of Polymeric Liquids, Vol.
  1}}}\ (\bibinfo  {publisher} {Wiley-Interscience, New York},\ \bibinfo {year}
  {1987})\ Chap.~\bibinfo {chapter} {3}\BibitemShut {NoStop}%
\bibitem [{\citenamefont {Larson}(1988)}]{Larson:88}%
  \BibitemOpen
  \bibfield  {author} {\bibinfo {author} {\bibfnamefont {R.~G.}\ \bibnamefont
  {Larson}},\ }\href@noop {} {\emph {\bibinfo {title} {Consitutive Equations
  for Polymer Melts and Solutions}}}\ (\bibinfo  {publisher} {Butterworths,
  Boston},\ \bibinfo {year} {1988})\BibitemShut {NoStop}%
\bibitem [{\citenamefont {Evans}\ \emph {et~al.}(2012)\citenamefont {Evans},
  \citenamefont {Zirkelbach}, \citenamefont {Schr\"oder-Turk}, \citenamefont
  {Kraynik},\ and\ \citenamefont {Mecke}}]{Evans:12}%
  \BibitemOpen
  \bibfield  {author} {\bibinfo {author} {\bibfnamefont {M.~E.}\ \bibnamefont
  {Evans}}, \bibinfo {author} {\bibfnamefont {J.}~\bibnamefont {Zirkelbach}},
  \bibinfo {author} {\bibfnamefont {G.~E.}\ \bibnamefont {Schr\"oder-Turk}},
  \bibinfo {author} {\bibfnamefont {A.~M.}\ \bibnamefont {Kraynik}}, \ and\
  \bibinfo {author} {\bibfnamefont {K.}~\bibnamefont {Mecke}},\ }\href
  {\doibase 10.1103/PhysRevE.85.061401} {\bibfield  {journal} {\bibinfo
  {journal} {Phys. Rev. E}\ }\textbf {\bibinfo {volume} {85}},\ \bibinfo
  {pages} {061401} (\bibinfo {year} {2012})}\BibitemShut {NoStop}%
\bibitem [{\citenamefont {Mecke}(1998)}]{Mecke:98}%
  \BibitemOpen
  \bibfield  {author} {\bibinfo {author} {\bibfnamefont {K.}~\bibnamefont
  {Mecke}},\ }\href@noop {} {\bibfield  {journal} {\bibinfo  {journal} {Int. J.
  Mod. Phys. B}\ }\textbf {\bibinfo {volume} {12}},\ \bibinfo {pages} {861}
  (\bibinfo {year} {1998})}\BibitemShut {NoStop}%
\bibitem [{\citenamefont {Arns}\ \emph {et~al.}(2001)\citenamefont {Arns},
  \citenamefont {Knackstedt}, \citenamefont {Pinczewski},\ and\ \citenamefont
  {Mecke}}]{Arns:01}%
  \BibitemOpen
  \bibfield  {author} {\bibinfo {author} {\bibfnamefont {C.~H.}\ \bibnamefont
  {Arns}}, \bibinfo {author} {\bibfnamefont {M.~A.}\ \bibnamefont
  {Knackstedt}}, \bibinfo {author} {\bibfnamefont {W.~V.}\ \bibnamefont
  {Pinczewski}}, \ and\ \bibinfo {author} {\bibfnamefont {K.}~\bibnamefont
  {Mecke}},\ }\href@noop {} {\bibfield  {journal} {\bibinfo  {journal} {Phys.
  Rev. E}\ }\textbf {\bibinfo {volume} {63}},\ \bibinfo {pages} {31112:1}
  (\bibinfo {year} {2001})}\BibitemShut {NoStop}%
\bibitem [{\citenamefont {Mickel}\ \emph {et~al.}(2008)\citenamefont {Mickel},
  \citenamefont {M\"unster}, \citenamefont {Jawerth}, \citenamefont {Vader},
  \citenamefont {Weitz}, \citenamefont {Sheppard}, \citenamefont {Mecke},
  \citenamefont {Fabry},\ and\ \citenamefont {Schr\"{o}der-Turk}}]{Mickel:08}%
  \BibitemOpen
  \bibfield  {author} {\bibinfo {author} {\bibfnamefont {W.}~\bibnamefont
  {Mickel}}, \bibinfo {author} {\bibfnamefont {S.}~\bibnamefont {M\"unster}},
  \bibinfo {author} {\bibfnamefont {L.~M.}\ \bibnamefont {Jawerth}}, \bibinfo
  {author} {\bibfnamefont {D.~A.}\ \bibnamefont {Vader}}, \bibinfo {author}
  {\bibfnamefont {D.~A.}\ \bibnamefont {Weitz}}, \bibinfo {author}
  {\bibfnamefont {A.~P.}\ \bibnamefont {Sheppard}}, \bibinfo {author}
  {\bibfnamefont {K.}~\bibnamefont {Mecke}}, \bibinfo {author} {\bibfnamefont
  {B.}~\bibnamefont {Fabry}}, \ and\ \bibinfo {author} {\bibfnamefont {G.~E.}\
  \bibnamefont {Schr\"{o}der-Turk}},\ }\href@noop {} {\bibfield  {journal}
  {\bibinfo  {journal} {Biophys. J.}\ }\textbf {\bibinfo {volume} {95}},\
  \bibinfo {pages} {6072} (\bibinfo {year} {2008})}\BibitemShut {NoStop}%
\bibitem [{\citenamefont {Liu}\ \emph {et~al.}(1995)\citenamefont {Liu},
  \citenamefont {Nagel}, \citenamefont {Schecter}, \citenamefont {Coppersmith},
  \citenamefont {Majumdar}, \citenamefont {Narayan},\ and\ \citenamefont
  {Witten}}]{Liu:95}%
  \BibitemOpen
  \bibfield  {author} {\bibinfo {author} {\bibfnamefont {C.~h.}\ \bibnamefont
  {Liu}}, \bibinfo {author} {\bibfnamefont {S.~R.}\ \bibnamefont {Nagel}},
  \bibinfo {author} {\bibfnamefont {D.~A.}\ \bibnamefont {Schecter}}, \bibinfo
  {author} {\bibfnamefont {S.~N.}\ \bibnamefont {Coppersmith}}, \bibinfo
  {author} {\bibfnamefont {S.}~\bibnamefont {Majumdar}}, \bibinfo {author}
  {\bibfnamefont {O.}~\bibnamefont {Narayan}}, \ and\ \bibinfo {author}
  {\bibfnamefont {T.~A.}\ \bibnamefont {Witten}},\ }\href {\doibase
  10.1126/science.269.5223.513} {\bibfield  {journal} {\bibinfo  {journal}
  {Science}\ }\textbf {\bibinfo {volume} {269}},\ \bibinfo {pages} {513}
  (\bibinfo {year} {1995})}\BibitemShut {NoStop}%
\bibitem [{\citenamefont {Cates}\ \emph {et~al.}(1998)\citenamefont {Cates},
  \citenamefont {Wittmer}, \citenamefont {Bouchaud},\ and\ \citenamefont
  {Claudin}}]{Cates:98}%
  \BibitemOpen
  \bibfield  {author} {\bibinfo {author} {\bibfnamefont {M.~E.}\ \bibnamefont
  {Cates}}, \bibinfo {author} {\bibfnamefont {J.~P.}\ \bibnamefont {Wittmer}},
  \bibinfo {author} {\bibfnamefont {J.-P.}\ \bibnamefont {Bouchaud}}, \ and\
  \bibinfo {author} {\bibfnamefont {P.}~\bibnamefont {Claudin}},\ }\href
  {\doibase 10.1103/PhysRevLett.81.1841} {\bibfield  {journal} {\bibinfo
  {journal} {Phys. Rev. Lett.}\ }\textbf {\bibinfo {volume} {81}},\ \bibinfo
  {pages} {1841} (\bibinfo {year} {1998})}\BibitemShut {NoStop}%
\bibitem [{\citenamefont {Sheppard}\ \emph {et~al.}(2005)\citenamefont
  {Sheppard}, \citenamefont {Sok},\ and\ \citenamefont
  {Averdunk}}]{Sheppard:05}%
  \BibitemOpen
  \bibfield  {author} {\bibinfo {author} {\bibfnamefont {A.~P.}\ \bibnamefont
  {Sheppard}}, \bibinfo {author} {\bibfnamefont {R.~M.}\ \bibnamefont {Sok}}, \
  and\ \bibinfo {author} {\bibfnamefont {H.}~\bibnamefont {Averdunk}},\
  }\href@noop {} {\bibfield  {journal} {\bibinfo  {journal} {In Proceedings of
  the International Symposium of the Society of Core Analysts, Toronto}\
  }\textbf {\bibinfo {volume} {paper SCA2005-20}},\ \bibinfo {pages} {21}
  (\bibinfo {year} {2005})}\BibitemShut {NoStop}%
\end{thebibliography}
\end{document}